\documentclass[twocolumn,english,notitlepage,twocolumn,superscriptaddress]{revtex4-1}
\usepackage[T1]{fontenc}
\usepackage[latin9]{inputenc}
\usepackage{verbatim}
\usepackage{mathrsfs}
\usepackage{amsmath}
\usepackage{graphicx}

\makeatletter

\usepackage{tikz}
\usetikzlibrary{calc}

\usepackage{babel}

\makeatother

\usepackage{babel}
\begin{document}
\title{Directional quantum random walk induced by coherence}
\author{Jin-Fu Chen}
\address{Beijing Computational Science Research Center, Beijing 100193, China}
\address{Graduate School of China Academy of Engineering Physics, No. 10 Xibeiwang
East Road, Haidian District, Beijing, 100193, China}
\author{Yu-Han Ma}
\email{yhma@csrc.ac.cn}

\address{Beijing Computational Science Research Center, Beijing 100193, China}
\address{Graduate School of China Academy of Engineering Physics, No. 10 Xibeiwang
East Road, Haidian District, Beijing, 100193, China}
\author{Chang-Pu Sun}
\email{cpsun@csrc.ac.cn}

\address{Beijing Computational Science Research Center, Beijing 100193, China}
\address{Graduate School of China Academy of Engineering Physics, No. 10 Xibeiwang
East Road, Haidian District, Beijing, 100193, China}
\date{\today}
\begin{abstract}
Quantum walk (QW), which is considered as the quantum counterpart
of the classical random walk (CRW), is actually the quantum extension
of CRW from the single-coin interpretation. The sequential unitary
evolution engenders correlation between different steps in QW and
leads to a ballistic position distribution. In this paper, we propose
an alternative quantum extension of CRW from the ensemble interpretation,
named quantum random walk (QRW), where the walker has many unrelated
coins, modeled as two-level systems, initially prepared in the same
state. We calculate the walker's position distribution in QRW for
different initial coin states with the coin operator chosen as Hadamard
matrix. In one-dimensional case, the walker's position is the asymmetric
binomial distribution. We further demonstrate that in QRW, coherence
leads the walker to perform directional movement. For an initially
decoherenced coin state, the walker's position distribution is exactly
the same as that of CRW. Moreover, we study QRW in 2D lattice, where
the coherence plays a more diversified role in the walker's position
distribution.
\end{abstract}
\maketitle

\section{Introduction}

In the classical random walk (CRW), the walker is usually assumed
to have one single coin. At each step, he flips the coin and decides
the moving direction according to the flipping result \citep{VANKAMPEN2007ix}.
The coin is either heads or tails after flipping, and then the walker
moves right or left accordingly in one-dimensional case. This is the
single-coin interpretation for CRW. Since the flipping process of
CRW eliminates the correlation between the coin and the walker, no
correlation exists between different steps. In other words, the coin
can be considered as independent coins for different steps. This indicates
that we can understand CRW with the \textsl{ensemble interpretation},
where the walker possesses many independent coins, and flips each
coin at each step.

It is conventionally understood that the quantum counterpart of CRW
is the quantum walk (QW) \citep{Ambainis_2001,KENDON_2007,Venegas_Andraca_2012}
(named as quantum random walk in early studies)\textit{,} the concept
of which was first proposed by Aharonov \citep{Aharonov_1993}. Different
from CRW, the walker's position distribution of QW is found to be
ballistic \citep{Abal_2006,Ermann_2006,Venegas_Andraca_2012}. QW
has been extensively studied to utilize its advantage in quantum computation
\citep{Shenvi_2003,Childs_2009,Lovett_2010}, quantum simulation \citep{Witthaut_2010,Mohseni_2008},
or to give a prototype to understand the quantum phase transition
and the topological phases \citep{Kitagawa_2012,Wang_2019}. Recently,
QW has been realized in experiment with different physical systems,
such as trapped atoms or ions \citep{Karski_2009,Z_hringer_2010,Schmitz_2009,Xue_2009},
optical systems \citep{Schreiber_2010,Broome_2010,Peruzzo_2010,Tang2018},
and superconducting qubit \citet{Yan2019}. Theoretical studies provide
the transition from QW to CRW in different fashions, with decoherence
approach \citep{Whitfield_2010,Ermann_2006,Brun_2003,Zhang2008},
or with random phase approach \citep{Ko_k_2006}. Actually, QW is
the quantum extension of CRW from the single-coin interpretation.
In the current version of QW, the state of the walker and the coin
is described by the quantum state in the corresponding Hilbert space
while the flipping process is considered as a unitary transform on
the coin \citep{Abal_2006,Venegas_Andraca_2012}. The unitary transform
engenders strong correlation between different steps. While in CRW,
the flipping process eliminates the correlation between the walker
and the coin, and every step is independent.

Inspired by the \textsl{ensemble interpretation} of CRW, we propose
a new quantum random walk (QRW) in this paper, where random means
each step is uncorrelated. QRW can be regarded as an alternative quantum
extension of CRW from the emsemble interpretation, while QW is the
quantum extension of CRW from the single-coin interpretation. In QRW,
the walker possesses many quantum coins, modeled as two-level systems
prepared in the same initial state. The walker flips each coin at
each step, described by a coin operator as a unitary evolution, and
moves according to the corresponding flipping result at each step.
Similar to QW \citep{Venegas_Andraca_2012}, the coin operator is
chosen as Hadamard matrix. We study the the walker's position distribution
in QRW with different coin's initial states. For an initially decoherenced
coin state, the  walker's position distribution recovers the result
of CRW. For an initial coin state with coherence, the walker's position
is shown to follows the asymmetric binomial distribution, where the
orientation of the walker is determined by the real part of the non-diagonal
term of the initial coin state.

This paper is organized as follows. In Sec. \ref{secII:From-classical-random},
we revisit CRW in the language of the density matrix and give the
two interpretation for CRW, the single-coin interpretation and the
\textit{ensemble interpretation}. In Sec. \ref{sec:Quantum-random-walk},
we propose QRW, as the quantum extension of CRW from the ensemble
interpretation, and discuss the walker's position distribution in
1D case. In Sec. \ref{sec:The-correlation-in}, we analyze the correlation
between different step in CRW, QW, and QRW. We thus clarify that it
is the difference in such correlation that makes the walker follows
different position distribution in those walk models. In Sec. \ref{subsec:2D-quantum-random},
we extend the framework of the new QRW to 2D lattice. Finally, the
conclusion of the main results is presented in Sec. \ref{sec:Conclusion-and-Discussion}.

\section{Revisit classical random walk with density matrix approach\label{secII:From-classical-random}}

\subsection{Single-coin interpretation}

As a preparation, we first revisit CRW in the language of the density
matrix. In the beginning, the position of the walker is set to the
origin of the coordinates $\left|0\right\rangle $, while the coin
stays at a mixed state
\begin{equation}
\rho_{c}=p_{1}^{(0)}\left|1\right\rangle _{c}\left\langle 1\right|+p_{-1}^{(0)}\left|-1\right\rangle _{c}\left\langle -1\right|.\label{eq:rouca}
\end{equation}
where $\left|1\right\rangle _{c}$ and $\left|-1\right\rangle _{c}$
represent the heads and tails of the coin respectively, with the corresponding
probability as $p_{1}^{(0)}$ and $p_{-1}^{(0)}=1-p_{1}^{(0)}$. The
non-diagonal term of the above density matrix is zero since the coin
is completely classical without any coherence in CRW. Such that, the
total initial state of the walker and the coin is
\begin{equation}
\rho(0)=\left|0\right\rangle _{w}\left\langle 0\right|\otimes\rho_{c}\label{eq:initial state}
\end{equation}

At each step, the walker flips the coin and moves according to the
flipping result
\begin{equation}
\rho(l+1)=\mathscr{T}\mathscr{C}\left[\rho(l)\right],\label{eq:rhok=00003DTCrhok-1}
\end{equation}
where $\rho(l)$ is the total density matrix of the walker and coin
after $k$-th step. Since the density matrix is always diagonal in
CRW, we can write $\rho(l)$ as 
\begin{equation}
\rho(l)=\sum_{x=-\infty}^{\infty}\sum_{u=\pm1}p_{x,u}(l)\left|x\right\rangle _{w}\left\langle x\right|\otimes\left|u\right\rangle _{c}\left\langle u\right|,\label{eq:densitymatrix-CRW}
\end{equation}
where $p_{x,u}(k)$ is the probability for the walker arrive at $x$
and the coin is at $\left|u\right\rangle _{c}$ state after $k$ step.
The flipping process $\mathscr{C}$ only operates on the coin, and
transforms the density matrix to $\tilde{\rho}(l)=\mathscr{C}\left[\rho(l)\right]$
as
\begin{equation}
\tilde{\rho}(l)=\sum_{x=-\infty}^{\infty}\sum_{u=\pm1}\tilde{p}_{x,u}(l)\left|x\right\rangle _{w}\left\langle x\right|\otimes\left|u\right\rangle _{c}\left\langle u\right|,
\end{equation}
with the new distribution $\tilde{p}_{x,u}(l)=\sum_{v=\pm1}p_{x,v}(l)p(u|v)$.
Here, $p(u|v)$ denotes the conditional probability for flipping the
coin from $\left|v\right\rangle _{c}$ state to $\left|u\right\rangle _{c}$
state with $v,u=\pm1$. For CRW, the state of the coin before and
after flipping should be independent, which requires the conditional
probability satisfies $p(u|1)=p(u|-1)$. After flipping, the new distribution
becomes 
\begin{equation}
\tilde{p}_{x,u}(l)=p_{x}(l)p_{u},\label{eq:recursion0}
\end{equation}
where $p_{x}(l)=p_{x,1}(l)+p_{x,-1}(l)$ gives the position distribution,
and $p_{u}=p(u|1)=p(u|-1)$ gives the coin distribution. It is clearly
seen in Eq. (\ref{eq:recursion0}) that the flipping process eliminates
the correlation between the walker and the coin. Therefore, the total
density matrix after flipping becomes a product state composed of
the walker and the coin as

\begin{equation}
\tilde{\rho}(l)=\left[\mathrm{Tr}_{c}\rho(l)\right]\otimes\tilde{\rho}_{c},\label{eq:reset}
\end{equation}
where the flipped coin state $\tilde{\rho}_{c}$ follows

\begin{equation}
\tilde{\rho}_{c}=p_{1}\left|1\right\rangle _{c}\left\langle 1\right|+p_{-1}\left|-1\right\rangle _{c}\left\langle -1\right|,\label{eq:C(r)}
\end{equation}
which is the same after flipping at different step. For CRW without
bias, all the conditional probabilities equals to $1/2$ and the flipped
coin state becomes the fully mixed state $\tilde{\rho}_{c}=1/2\left(\left|1\right\rangle _{c}\left\langle 1\right|+\left|-1\right\rangle _{c}\left\langle -1\right|\right)$.

After flipping, the walker moves according to the flipped coin state
$\tilde{\rho}_{c}$ through the transition process $\mathscr{T}\left[\tilde{\rho}(l)\right]=T\tilde{\rho}(l)T^{\dagger}$
with the transition operator
\begin{equation}
T=\sum_{x=-\infty}^{\infty}\sum_{u=\pm1}\left|x+u\right\rangle _{w}\left\langle x\right|\otimes\left|u\right\rangle _{c}\left\langle u\right|,\label{eq:T-onecoin}
\end{equation}
which means the walker moves right (left) when the coin stays at $\left|1\right\rangle _{c}$
($\left|-1\right\rangle _{c}$). Thus, after $(l+1)$-th step, the
total density matrix $\rho(l+1)=\mathscr{T}\left[\tilde{\rho}(l)\right]$
is explicitly obtained as
\begin{equation}
\rho(l+1)=\sum_{x=-\infty}^{\infty}\sum_{u=\pm1}\tilde{p}_{x-u,u}(l)\left|x\right\rangle _{w}\left\langle x\right|\otimes\left|u\right\rangle _{c}\left\langle u\right|.
\end{equation}
Together with Eq. (\ref{eq:densitymatrix-CRW}), we obtain the recursion
relation 
\begin{align}
p_{x,u}(l+1) & =\tilde{p}_{x-u,u}(l),\,u=\pm1.\label{eq:recursion}
\end{align}
We remark that the transition process is a unitary evolution and remains
the same as QW.

According to the recursion relations of Eqs. (\ref{eq:recursion0})
and (\ref{eq:recursion}), it follows from Eq. (\ref{eq:rhok=00003DTCrhok-1})
that the total density matrix of the walker and coin after $n$ steps
is
\begin{widetext}
\begin{equation}
\rho(n)=\sum_{j=0}^{n}\left|2j-n\right\rangle _{w}\left\langle 2j-n\right|\otimes p_{1}^{j}p_{-1}^{n-j}\left[\left(\begin{array}{c}
n-1\\
j-1
\end{array}\right)\left|1\right\rangle _{c}\left\langle 1\right|+\left(\begin{array}{c}
n-1\\
j
\end{array}\right)\left|-1\right\rangle _{c}\left\langle -1\right|\right],\label{eq:CRW density one coin}
\end{equation}
\end{widetext}

By tracing over the coin's degree of freedom in $\rho(n)$, we obtain
the probability for the walker arriving at the position $2j-n$ after
$n$ steps as 
\begin{equation}
P_{2j-n}(n)=\left(\begin{array}{c}
n\\
j
\end{array}\right)p_{1}^{j}p_{-1}^{n-j}.\label{eq:position distribution-CRW}
\end{equation}
This probability distribution, known as the binomial distribution,
describes the walker's position distribution in the classical random
walk. The expectation and variance of the walker's position \citep{VANKAMPEN2007ix}
are given by

\begin{equation}
\left\langle x\right\rangle =n(p_{1}-p_{-1}),\label{eq:xexpectation}
\end{equation}
and

\begin{equation}
\left\langle \Delta x^{2}\right\rangle =\left\langle x^{2}\right\rangle -\left\langle x\right\rangle ^{2}=4np_{1}p_{-1},\label{eq:xvariance}
\end{equation}
respectively. When $p_{1}=p_{-1}=1/2$, the binomial distribution
of Eq. (\ref{eq:position distribution-CRW}) is symmetric. In this
case, the expected position of the walker after $n$ steps is just
the origin of the coordinates, which can be easily cheeked from Eq.
(\ref{eq:xexpectation}). Otherwise, for $p_{1}\neq p_{-1}$, the
position distribution is asymmetric, the walker will thus perform
directional walking, i.e., $\left\langle x\right\rangle \neq0$, and
the CRW is directional.

\subsection{Ensemble interpretation with many coins\label{subsec:Ensemble-interpretation-with}}

In the above discussion, the flipping process $\mathscr{C}$ eliminates
the correlation between the coin and the the walker at every step,
and the flipped coin state $\tilde{\rho}_{c}$ does not depend on
the previous state $\rho(k)$. The coin can be considered as independent
coins for different steps. This is the \textsl{ensemble interpretation}
for CRW. In the following discussion, we will obtain the same result
of the position distribution based on the ensemble interpretation.
Suppose the walker possesses many coins, the number of which equals
the total step number $n$ the walker will move. The total Hilbert
space is the product of the walker space and the space for each coin
\begin{equation}
\mathscr{H}_{\mathrm{T}}=\mathscr{H}_{w}\otimes\bigotimes_{l=1}^{n}\mathscr{H}_{l,c}.
\end{equation}
Ar the beginning, all the coins satisfy the same distribution $\rho_{c}$
by Eq. (\ref{eq:rouca}). Now, the initial density matrix of the walker
and all the coins reads

\begin{equation}
\rho(0)=\left|0\right\rangle _{w}\left\langle 0\right|\otimes\bigotimes_{l=1}^{n}\rho_{l,c},\label{eq:rou(0)}
\end{equation}
where $l$ distinguishes different coins. At the $l$-th step, the
walker flips the $l$-th coin and moves according to the flipping
result, namely,
\begin{equation}
\rho(l)=\mathscr{T}_{l}\mathscr{C}_{l}\left[\rho(l-1)\right],\label{eq:11}
\end{equation}
The flipping process $\mathscr{C}_{l}$ transforms the $l$-th coin's
state from $\rho_{l,c}$ to $\tilde{\rho}_{l,c}$, where $\tilde{\rho}_{l,c}$
follows the same form as Eq. (\ref{eq:C(r)}). And the transition
process 
\begin{equation}
\mathscr{T}_{l}\left(\rho\right)=T_{l}\rho T_{l}^{\dagger}\label{eq:QRW transition process}
\end{equation}
is realized with the transition operator

\begin{equation}
T_{l}=\sum_{x=-\infty}^{\infty}\sum_{u=\pm1}\left|x+u\right\rangle _{w}\left\langle x\right|\otimes\left|u\right\rangle _{l}\left\langle u\right|\otimes\bigotimes_{j\ne l}^{n}I_{j}.\label{eq:T-1D-1}
\end{equation}
Here, $I_{j}$ is the $2\times2$ identity matrix for the $j$-th
coin. So that, the total density matrix after $n$ step is $\rho\left(n\right)=\prod_{l=1}^{n}\left(\mathscr{T}_{l}\mathscr{C}_{l}\right)\left[\rho(0)\right]$,
which can be explicitly written as
\begin{equation}
\rho(n)=\sum_{\{u_{l}\}}\left|\sum_{l}u_{l}\right\rangle _{w}\left\langle \sum_{l}u_{l}\right|\otimes\bigotimes_{l=1}^{n}\left(p_{u_{l}}\left|u_{l}\right\rangle _{l}\left\langle u_{l}\right|\right).\label{eq:13}
\end{equation}
In the summation, $u_{l}=\pm1$ gives the direction for each step.
By tracing over the space of all the coins, the probability for the
walker arriving at the position $x$ after $n$ steps is obtained
as 
\begin{equation}
P_{x}(n)=\sum_{\left\{ u_{l}\right\} :\sum_{l}u_{l}=x}\prod_{l=1}^{n}p_{u_{l}}.
\end{equation}
The limitation on the path $\sum_{l=1}^{n}u_{l}=x$ requires $(x+n)/2$
right steps and $(x-n)/2$ left steps along $n$ steps, and $(x\pm n)/2$
needs to be a positive integer otherwise the probability is zero.
Then the probability at the position $x$ is obtained explicitly as

\begin{equation}
P_{x}(n)=\begin{cases}
\left(\begin{array}{c}
n\\
\frac{n+x}{2}
\end{array}\right)p_{1}^{\frac{n+x}{2}}p_{-1}^{\frac{n-x}{2}}. & n+x\mathrm{\:is\:even}\\
0 & n+x\mathrm{\:is\:odd}
\end{cases}.\label{eq:15-1}
\end{equation}

It is clearly seen from Eq. (\ref{eq:15-1}) that the walker's position
distribution is exactly the same as that of Eq. (\ref{eq:CRW density one coin})
in the one-coin case by setting $j=(n+x)/2$. Therefore, the equivalence
between the one-coin interpretation and many-coin interpretation (ensemble
interpretation) for CRW is proved. In further investigation below,
we will extend the ensemble interpretation to quantum random walk
to study the effect of the initial coherence of the coin.

\section{Quantum random walk\label{sec:Quantum-random-walk}}

In this section we will discuss the quantum random walk (QRW) in one-dimensional
space from the perspective of the ensemble interpretation of Sec.
\ref{subsec:Ensemble-interpretation-with}. For QRW, the total initial
density matrix of the system is also described by Eq. (\ref{eq:rou(0)}),
where the initial state of the $l$-th coin is now assumed to be 
\begin{table}
\includegraphics[width=8.5cm]{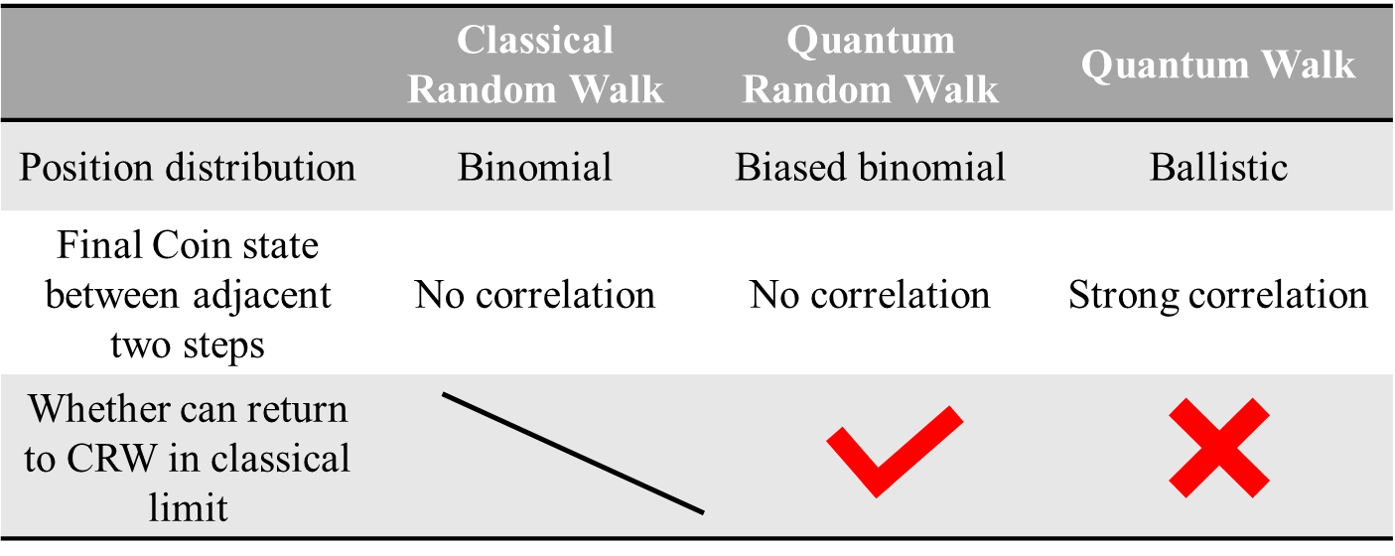}

\caption{\label{table:The-relation-between-1}The relation between classical
random walk (CRW), quantum walk (QW) and quantum random walk (QRW)
defined in this paper. The explicit position distribution of the walker
in CRW, and QRW is given by Eqs. (\ref{eq:position distribution-CRW})
and (\ref{eq:29-1}) respectively. The ballistic distribution of QW
is shown by the green circle markered line in Fig. \ref{fig:Comparing-the-position}.
Detailed discussion about the correlation between different steps
is demonstrated in Sec. \ref{sec:The-correlation-in}.}
\end{table}

\begin{equation}
\rho_{l,c}=\left(\begin{array}{cc}
p_{1}^{(0)} & \eta\\
\eta^{*} & p_{-1}^{(0)}
\end{array}\right),\label{eq:16-1}
\end{equation}
where the non-diagonal term $\eta$ characterizes the coherence exists
in the coin state. We consider a unitary flipping process $\mathscr{C}_{l}$
at $l$-th step acting on the $l$-th coin
\begin{equation}
\mathscr{C}_{l}\left(\rho_{l,c}\right)=C_{l}\rho_{l,c}C_{l}^{\dagger}\equiv\tilde{\rho}_{l,c},\label{eq:QRW flipping process}
\end{equation}
where $\tilde{\rho}_{l,c}$ is called the flipped state of the coin.
The coin operator only acts on the $l$-th coin

\begin{equation}
C_{l}=I_{w}\otimes\bigotimes_{j\ne l}^{n}I_{j}\otimes\tilde{C}_{l},\label{eq:Ck}
\end{equation}
where $\tilde{C}_{l}$ is a U(2) matrix for the $l$-th coin, and
$I_{w}$ is the identity matrix in the walker's Hilbert space. For
a general SU(2) matrix 
\begin{equation}
\tilde{C}_{l}=\left(\begin{array}{cc}
a & b\\
-b^{*} & a^{*}
\end{array}\right),
\end{equation}
it follows from Eqs. (\ref{eq:QRW flipping process}) and (\ref{eq:Ck})
that the coin state after flipping becomes

\begin{equation}
\tilde{\rho}_{l,c}=\sum_{u_{l},v_{l}}\rho_{u_{l}v_{l}}\left|u_{l}\right\rangle _{l}\left\langle v_{l}\right|,\label{eq:C(rou)}
\end{equation}
where $u_{l},v_{l}=\pm1$ and

\begin{align}
\begin{aligned}\rho_{1,1} & =p_{1}^{(0)}\left|a\right|^{2}+\eta^{*}a^{*}b+\eta ab^{*}+p_{-1}^{(0)}\left|b\right|^{2}\\
\rho_{1,-1} & =a\left(a\eta+bp_{-1}^{(0)}\right)-b\left(ap_{1}^{(0)}+b\eta^{*}\right)\\
\rho_{-1,1} & =a^{*}\left(a^{*}\eta^{*}+b^{*}p_{2}^{(0)}\right)-b^{*}\left(a^{*}p_{1}^{(0)}+b^{*}\eta\right)\\
\rho_{-1,-1} & =p_{-1}^{(0)}\left|a\right|^{2}-\eta^{*}a^{*}b-\eta ab^{*}+p_{1}^{(0)}\left|b\right|^{2}.
\end{aligned}
\label{eq:p11-p-1-1}
\end{align}

According to Eq. (\ref{eq:11}), the total density matrix after $n$-th
step is

\begin{equation}
\rho(n)=\left(\prod_{l=1}^{n}T_{l}C_{l}\right)\rho(0)\left(\prod_{l=1}^{n}T_{l}C_{l}\right)^{\dagger}.\label{eq:17}
\end{equation}
Since $\left[C_{l},T_{l^{\prime}}\right]=0$ commutes for different
step $l\neq l^{\prime}$, we first act all the coin operators on the
initial state of the coins

\begin{equation}
\prod_{l=1}^{n}C_{l}\rho_{l,c}C_{l}^{\dagger}=\prod_{l=1}^{n}\tilde{\rho}_{l,c}.
\end{equation}
Then, Eq. (\ref{eq:17}) is rewritten as

\begin{equation}
\rho(n)=\left(\prod_{l=1}^{n}T_{l}\right)\left(\left|0\right\rangle \left\langle 0\right|\otimes\bigotimes_{l^{\prime}=1}^{n}\tilde{\rho}_{l^{\prime},c}\right)\left(\prod_{l=1}^{n}T_{l}\right)^{\dagger}.\label{eq:rou(n)}
\end{equation}
Substituting Eq. (\ref{eq:T-1D-1}) into Eq. (\ref{eq:rou(n)}), we
obtain 

\begin{equation}
\rho(n)=\sum_{\left\{ u_{l},v_{l}\right\} }\left|\sum_{l=1}^{n}u_{l}\right\rangle _{w}\left\langle \sum_{l=1}^{n}v_{l}\right|\otimes\bigotimes_{l=1}^{n}\rho_{u_{l}v_{l}}\left|u_{l}\right\rangle _{l}\left\langle v_{l}\right|.\label{eq:density k step}
\end{equation}
The position distribution of the walker is determined by the diagonal
elements of the density matrix of the flipped coin $\tilde{\rho}_{l,c}$.
The probability at the position $x$ after $n$ steps $P_{x}(n)=\mathrm{\mathrm{Tr}}_{c}\left\langle x\right|\rho(n)\left|x\right\rangle _{w}$
is obtained from Eq. (\ref{eq:density k step}) by tracing over the
freedom of the coins as
\begin{equation}
P_{x}(n)=\left(\begin{array}{c}
n\\
\frac{n+x}{2}
\end{array}\right)\rho_{1,1}^{\frac{n+x}{2}}\rho_{-1,-1}^{\frac{n-x}{2}},\label{pxn-QRW}
\end{equation}
where the corresponding transition probabilities $\rho_{1,1}$ and
$\rho_{-1,-1}$ are given in Eq. (\ref{eq:p11-p-1-1}). The walker's
position distribution by Eq. (\ref{pxn-QRW}) for QRW is a binomial
distribution with the probabilities $\rho_{1,1},\rho_{-1,-1}$, the
same as the distribution of a directional CRW. In QRW, the walker
flips different coins at different steps, hence each step is independent.
While in QW, the position distribution is shown to be ballistic distribution,
which strongly depends on the initial coin state \citep{Venegas_Andraca_2012}.
The non-binomial distribution comes from the strong correlation between
different steps, which will be specifically discussed in Sec. \ref{sec:The-correlation-in}.
To briefly show the similarities and differences between CRW, QW and
QRW, we illustrate their typical characteristics in Tab. \ref{table:The-relation-between-1}.

In order to understand the origin of the bias in QRW, we need to figure
out what determines the transition probabilities $\rho_{1,1}$ and
$\rho_{-1,-1}$. For the coin operator chosen as the Hadamard matrix
\begin{equation}
\tilde{C}_{l}=\frac{1}{\sqrt{2}}\left(\begin{array}{cc}
1 & 1\\
1 & -1
\end{array}\right),\label{ck-QRW}
\end{equation}
the transition probabilities follow as $\rho_{1,1}=1/2+\mathrm{Re}\eta$,
and $\rho_{-1,-1}=1/2-\mathrm{Re}\eta$. Therefore, after $n$ steps,
the position distribution of the walker is given by Eq. (\ref{pxn-QRW})
as 
\begin{equation}
P_{x}(n)=\left(\begin{array}{c}
n\\
\frac{n+x}{2}
\end{array}\right)(\frac{1}{2}+\mathrm{Re}\eta)^{\frac{n+x}{2}}(\frac{1}{2}-\mathrm{Re}\eta)^{\frac{n-x}{2}},\label{eq:29-1}
\end{equation}
which indicates that the bias is only determined by the real part
of the non-diagonal term. When the real part of the non-diagonal term
in the coin's density matrix is zero, i.e. $\mathrm{Re}\eta=0$, the
result returns back to  CRW without bias.

\begin{figure}
\begin{centering}
\includegraphics[width=8.5cm]{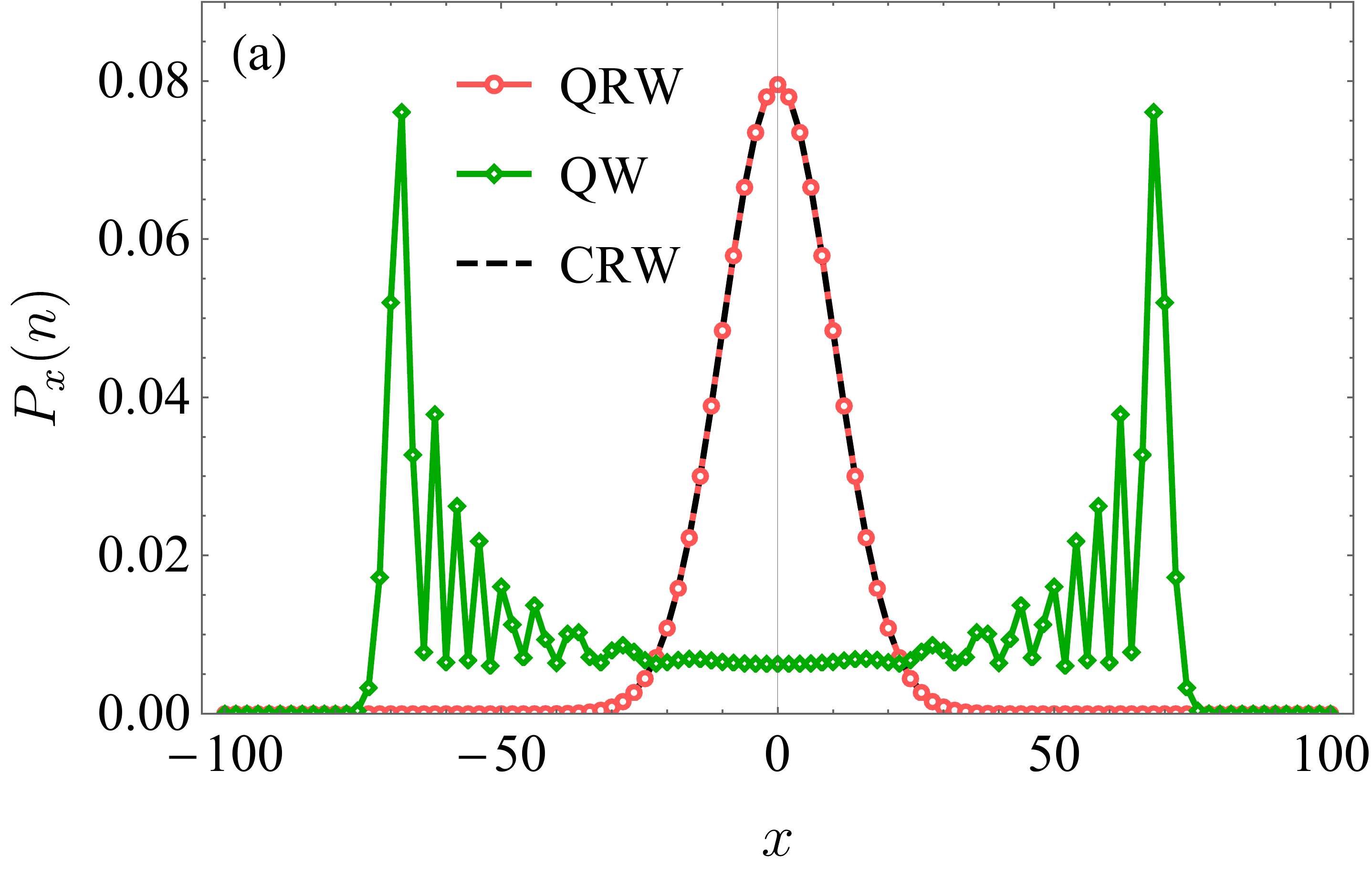}
\par\end{centering}
\centering{}\includegraphics[width=8.5cm]{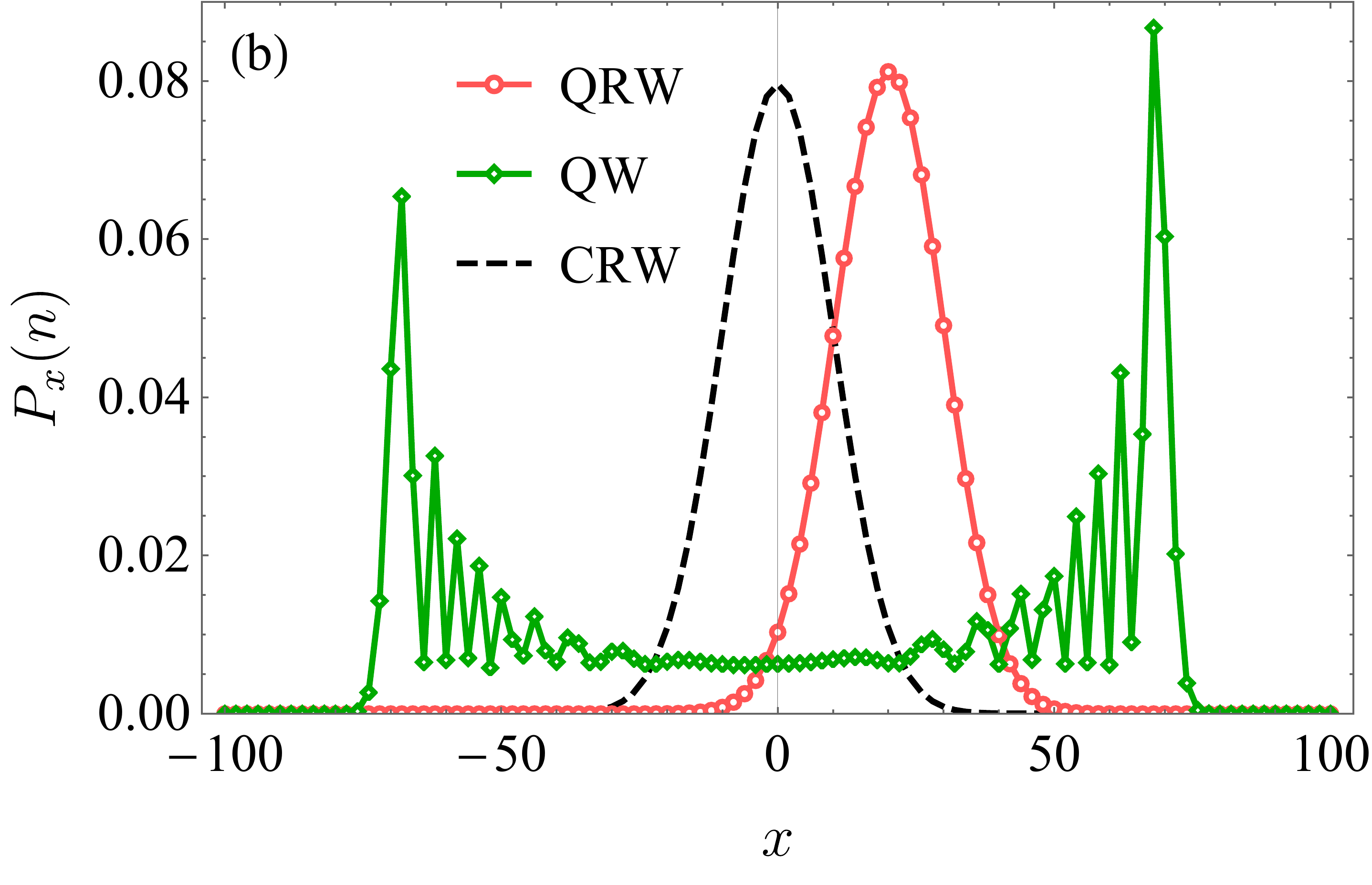}\caption{\label{fig:Comparing-the-position}The walker's position distribution
$P_{x}(n)$ as the function of position $x$ for quantum random walk
(red square markered line), quantum walk (green circle markered line),
classical  walk (black-dashed line). The total step number is chosen
as $n=100$ , and we only plot the probability at the even lattice
since the probability at the odd lattice is zero. The classical random
walk is given as an symmetric binomial distribution, which follows
from Eq. (\ref{eq:position distribution-CRW}) with $p_{1}=p_{-1}=0.5$.
(a) The coin initial state is chosen as $p_{1}=p_{-1}=0.5,\,\eta=0$.
The walker's position distribution of  the quantum random walk returns
to the one of the classical random walk. (b) The coin's initial state
is chosen as $p_{1}=p_{-1}=0.5,\,\eta=0.1$. The initial coherence
of the coin leads to asymmetric binomial distribution for the quantum
random walk.}
\end{figure}

As a comparison, we demonstrate the position distribution of QRW,
QW and CRW in Fig. \ref{fig:Comparing-the-position}. The total step
number is $n=100$, and we only plot the probability at the even lattice
since the probability at the odd lattice is zero. In Fig. \ref{fig:Comparing-the-position}(a),
we consider an initially decoherenced coin state with $p_{1}=p_{-1}=0.5,\,\eta=0$.
The position distribution of QRW (red square marked line) returns
to the one of CRW (black dashed line) while the position distribution
of QW is ballistic (green circle marked line). In Fig. \ref{fig:Comparing-the-position}(b),
we choose the initial state with coherence by setting $p_{1}=p_{-1}=0.5,\,\eta=0.1$.
The positive non-diagonal term of the density matrix results in the
right-hand movement for the QRW, namely, the coherence of the coin
induces the asymmetry in the corresponding position distribution.

With the position distribution given by Eq. (\ref{eq:29-1}), we obtain
the expectation and the variance of the walker's position after $n$
steps, according to Eqs. (\ref{eq:xexpectation}) and (\ref{eq:xvariance}),
as
\begin{equation}
\left\langle x\right\rangle =2n\mathrm{Re}\eta,\label{eq:xbar}
\end{equation}
and
\begin{equation}
\left\langle \Delta x^{2}\right\rangle =n\left[1-\left(2\mathrm{Re}\eta\right)^{2}\right],\label{x^2 bar}
\end{equation}
respectively. The above two relations of Eqs. (\ref{eq:xbar}) and
(\ref{x^2 bar}) are the main results of this paper, which show that
the coherence in the initial coin state results in the directional
moving of the walker.

\section{The correlation in quantum walk\label{sec:The-correlation-in}}

In Sec. \ref{secII:From-classical-random} and \ref{sec:Quantum-random-walk},
we have discussed the position distribution in CRW and QRW, and demonstrate
that no correlation exists in CRW and QRW between different steps.
In this section, we will show that the correlation indeed exists between
different steps in QW, which is qualified through the convariance
of the coin state between the initial time and final time.

In QW, the walker has only one coin, and the one-step evolution is
described by Eq. (\ref{eq:rhok=00003DTCrhok-1}) with the same transition
process as in CRW. Different from CRW, the flipping process in QW
is substituted by a unitary evolution $\mathscr{C}\left(\rho\right)=C\rho C^{\dagger}$
with the Hadamard matrix 
\begin{equation}
C=\frac{1}{\sqrt{2}}\left(\begin{array}{cc}
1 & 1\\
1 & -1
\end{array}\right)\label{eq:C one coin}
\end{equation}
 operating on the coin state.

The density matrix of the walker and coin after $n$ steps follows
from Eq. \ref{eq:rhok=00003DTCrhok-1} as
\begin{equation}
\rho(n)=(TC)^{n}\rho(0)(TC)^{\dagger n}.
\end{equation}
The transition operator $T$ is defined in Eq. (\ref{eq:T-onecoin}),
and the initial state $\rho(0)$ is given as Eq. (\ref{eq:initial state}).
To describe the coin's distribution after $n$ steps, we perform a
measurement of the Pauli operator $\sigma_{z}$ on the coin, as $\sigma_{z}\left|\pm1\right\rangle _{c}=\pm\left|\pm1\right\rangle _{c}$.
The expectation of $\sigma_{z}$ after $n$ steps is 
\begin{equation}
\left\langle \sigma_{z}(n)\right\rangle =\mathrm{Tr}\left[\sigma_{z}\rho_{c}(n)\right],
\end{equation}
where $\rho_{c}(n)$ is the reduced density matrix of the coin after
$n$ steps.
\begin{widetext}
To figure out the correlation between the initial time and final time,
we perform a joint measurement of $\sigma_{z}$ at the initial time.and
after $n$ steps, the expectation $\left\langle \sigma_{z}(n)\sigma_{z}(0)\right\rangle $
follows 
\begin{equation}
\left\langle \sigma_{z}(n)\sigma_{z}(0)\right\rangle =\sum_{\nu_{1},\nu_{2}=\pm1}\nu_{1}\nu_{2}\mathrm{Tr}\left[E_{\nu_{2}}U(n)E_{\nu_{1}}\rho(0)E_{\nu_{1}}U^{\dagger}(n)E_{\nu_{2}}\right],\label{convariance}
\end{equation}
where $\nu_{1}$($\nu_{2}$) gives the measurement result at the initial
time (after $n$ steps), and $E_{\nu_{\alpha}}=\left|\nu_{\alpha}\right\rangle _{c}\left\langle \nu_{\alpha}\right|,\alpha=1,2$
is the projection operator. Note that the non-diagonal term of $\rho(0)$
vanishes in the initial measurement, i.e.

\begin{equation}
E_{\nu_{1}}\rho(0)E_{\nu_{1}}=\left|0\right\rangle _{w}\left\langle 0\right|\otimes p_{\nu_{1}}^{(0)}\left|\nu_{1}\right\rangle _{c}\left\langle \nu_{1}\right|.
\end{equation}
Hence, we only need to consider for the diagonal initial state without
coherence. The correlation in the coin is described by the covariance

\begin{equation}
\left\langle \Delta\left[\sigma_{z}(n)\sigma_{z}(0)\right]\right\rangle =\left\langle \sigma_{z}(n)\sigma_{z}(0)\right\rangle -\left\langle \sigma_{z}(n)\right\rangle \left\langle \sigma_{z}(0)\right\rangle ,\label{eq:covirance define}
\end{equation}
which is obtained explicitly as (detailed derivation  in Appendix
\ref{sec:The-Covariance-for})
\begin{equation}
\left\langle \Delta\left[\sigma_{z}(n)\sigma_{z}(0)\right]\right\rangle =\left(1-\frac{\sqrt{2}}{2}+\frac{\left(-1\right)^{n}}{2\pi}\int_{-\pi}^{\pi}\left(\frac{\cos\left(2\omega_{k}n\right)}{1+\left(\cos k\right)^{2}}\right)\mathrm{d}k\right)\left[1-\left(p_{1}^{(0)}-p_{-1}^{(0)}\right)^{2}\right],\label{eq:covariance exact}
\end{equation}
with $\omega_{k}=\arcsin\left(\sin k/\sqrt{2}\right).$ In the large
$n$ limit ($n\rightarrow\infty$), the integral in Eq. (\ref{eq:covariance exact})
diminishes due to the highly oscillated term $\cos\left(2\omega_{k}n\right)$,
so that the covariance approaches a constant
\begin{equation}
\lim_{n\rightarrow\infty}\left\langle \Delta\left[\sigma_{z}(n)\sigma_{z}(0)\right]\right\rangle =\left(1-\frac{\sqrt{2}}{2}\right)\left[1-\left(p_{1}^{(0)}-p_{-1}^{(0)}\right)^{2}\right].\label{eq:covariance}
\end{equation}
The non-zero covariance suggests that the correlation generates through
the coin's flipping process, where the final coin state is correlated
to the initial coin state.
\end{widetext}

\begin{figure}
\includegraphics[width=8.5cm]{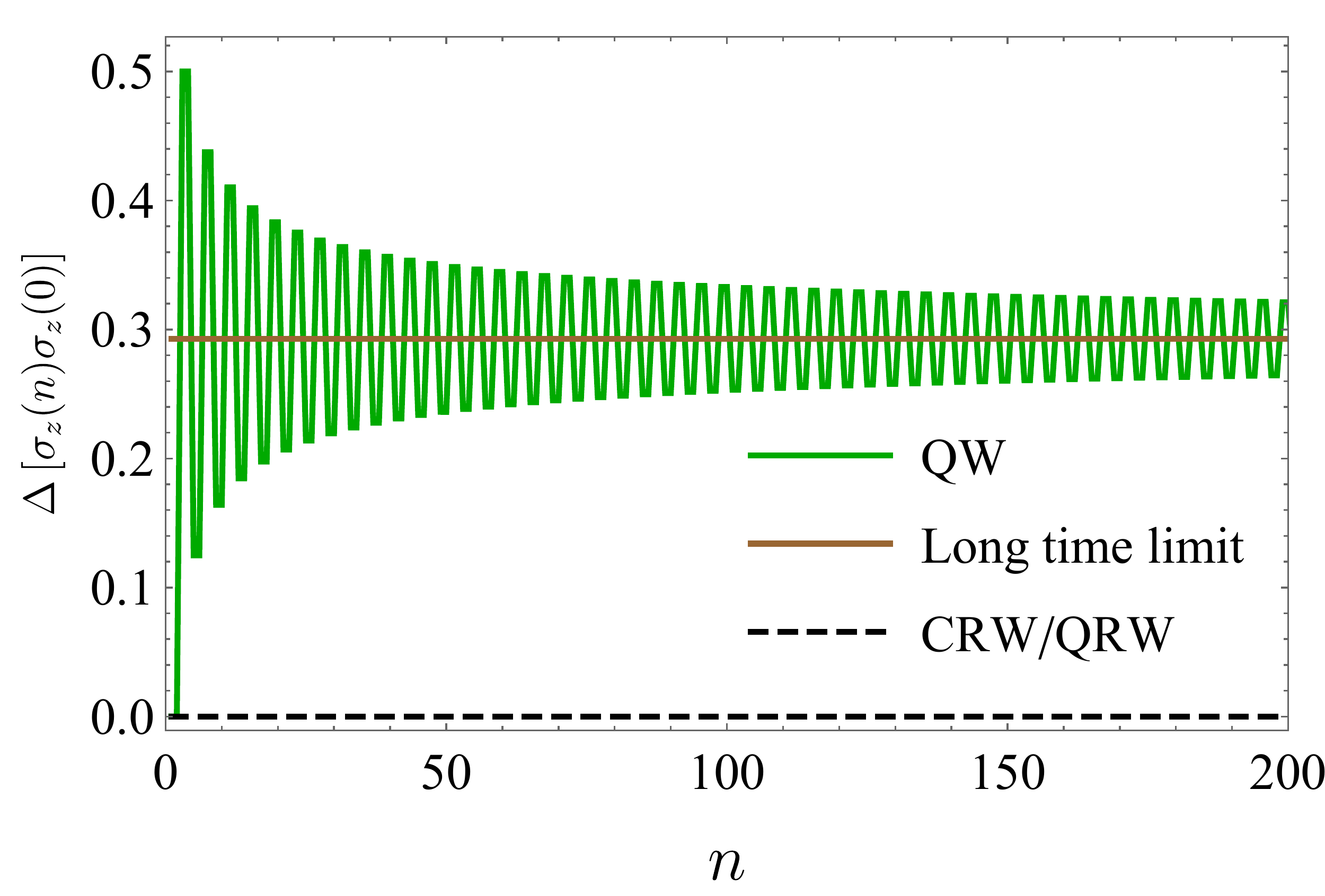}

\caption{(a) The covariance $\left\langle \Delta\left[\sigma_{z}(n)\sigma_{z}(0)\right]\right\rangle $
varies with different steps. The initial coin density matrix is chosen
as the maximal mixed state, and the coin operator is chosen as the
Hadamard matrix. The Green curve gives the covariance for quantum
walk by Eq. (\ref{eq:covariance exact}), while the horizontal dashed
line gives the covariance for classical random walk and quantum random
walk. The brown horizontal solid line shows the covariance approaches
to the constant $1-\sqrt{2}/2$, as predicted by Eq. (\ref{eq:covariance})
at large $n$ limit.}

\label{figprobabilityatheats}
\end{figure}

In Fig. \ref{figprobabilityatheats}, we illustrate the covariance
$\left\langle \Delta\left[\sigma_{z}(n)\sigma_{z}(0)\right]\right\rangle $
of QW, QRW, and CRW, where the initial coin is chosen as the maximally
mixed state with $p_{1}^{(0)}=p_{-1}^{(0)}=0.5$. It is clearly seen
in Fig. \ref{figprobabilityatheats} that the covariance of QW (Green
curve), given by Eq. (\ref{eq:covariance exact}), oscillates with
the increasing of $n$, and gradually converges to a non-zero constant
$1-1/\sqrt{2}$ (the brown horizontal line), which is consistent with
the analytical result of Eq. (\ref{eq:covariance}).

The non-zero covariance implies the coin state after $n$ steps is
correlated to the initial coin state, as shown in Tab. \ref{table:The-relation-between-1},
which implies that the coin will remember its initial state for no
matter how many steps the walker moves. The covariance of CRW and
QRW is both zero (the dashed horizontal line). In CRW, we have $\left\langle \sigma_{z}(n)\sigma_{z}(0)\right\rangle =\left\langle \sigma_{z}(n)\right\rangle \left\langle \sigma_{z}(0)\right\rangle $
which leads the corresponding covariance equals zero, and indicates
the flipping process at each step is independent. In QRW, the flipping
process for different coins at different steps is independent, and
thus the covariance of QRW $\left\langle \Delta\sigma_{z}^{n}(n)\sigma_{z}^{1}(0)\right\rangle =\left\langle \sigma_{z}^{n}(n)\sigma_{z}^{1}(0)\right\rangle -\left\langle \sigma_{z}^{n}(n)\right\rangle \left\langle \sigma_{z}^{1}(0)\right\rangle $
is also zero, where $\sigma_{z}^{1}(0)$ measures the state for the
first coin before walking, and $\sigma_{z}^{n}(n)$ measures the state
of the $n$-th coin after $n$ steps. Therefore, we state that no
correlation exists between different steps in CRW and QRW, while strong
correlation exists in QW.

\section{quantum random walk in 2D lattice\label{subsec:2D-quantum-random}}

With the theoretical framework of one-dimensional quantum random walk
established in Sec. \ref{sec:Quantum-random-walk}, it is convenient
for us to discuss QRW in two-dimensional lattice. Interestingly, unlike
the 1D QRW, in the 2D case, the influence of the coherence in the
coin's initial state on the position distribution of the the walker
is more complicated, as demonstrated in this section.

For QRW in 2D lattice, the Hilbert space of the walker is expanded
as $\left\{ \left|\vec{r}\right\rangle |\,\vec{r}=\left(x,y\right),x,y\in Z\right\} $.
And the Hilbert for each coin space is four dimension $\left\{ \left|\vec{u}\right\rangle \left|\text{\ensuremath{\vec{u}}=\ensuremath{\vec{R},\,\vec{L},\,\vec{U},\,\vec{D}}}\right.\right\} $,
to determine the walker moves right $\vec{R}=(1,0)$, left $\vec{L}=(-1,0)$,
up $\vec{U}=(0,1)$, and down $\vec{D}=(0,-1)$ correspondingly. We
still consider the walker initially stays at the origin of the coordinates
$\left|(0,0)\right\rangle _{w}$, and has many coins prepared in the
same initial state. In this situation, the initial density matrix
for the walker and the coins follows

\begin{equation}
\rho(0)=\left|(0,0)\right\rangle _{w}\left\langle (0,0)\right|\otimes\bigotimes_{l=1}^{n}\rho_{l,c},
\end{equation}
where $\rho_{l,c}$ is the density matrix of the $l$-th coin, and
can be represented by a general non-negative $4\times4$ Hermite matrix
as
\begin{equation}
\rho_{k,c}=\left(\begin{array}{cccc}
q_{1} & \eta_{12} & \eta_{13} & \eta_{14}\\
\eta_{21} & q_{2} & \eta_{23} & \eta_{24}\\
\eta_{31} & \eta_{32} & q_{3} & \eta_{34}\\
\eta_{41} & \eta_{42} & \eta_{43} & q_{4}
\end{array}\right),\label{eq:rou4*4}
\end{equation}
The coin flipping process and the transition process are the same
as that described by Eq. (\ref{eq:QRW flipping process}) and Eq.
(\ref{eq:T-1D-1}). The coin operator $C_{l}$ is also given by Eq.
(\ref{eq:Ck}), where $\tilde{C}_{l}$ can be chosen as a general
U(4) matrix. We consider the Grover coin acting on the $l$-th coin
\citep{Grover_1997,Shenvi_2003}, i.e. 
\begin{equation}
\tilde{C}_{l}=\frac{1}{2}\left(\begin{array}{cccc}
-1 & 1 & 1 & 1\\
1 & -1 & 1 & 1\\
1 & 1 & -1 & 1\\
1 & 1 & 1 & -1
\end{array}\right).
\end{equation}
Analogous to Eq. (\ref{eq:T-1D-1}) in the 1D case, the position of
walker changes according to the state of the $l$-th coin $\left|\vec{u}\right\rangle _{l}$
with $u=R,L,U,D$. So that, the transition operator for the 2D case
reads

\begin{equation}
T_{l}=\sum_{\vec{u}}\sum_{\vec{r}}\left|\vec{r}+\vec{u}\right\rangle _{w}\left\langle \vec{r}\right|\otimes\left|\vec{u}\right\rangle _{l}\left\langle \vec{u}\right|\otimes\bigotimes_{j\ne l}^{n}I_{j},
\end{equation}
Similar to the 1D QRW, the position distribution is unique determined
by the diagonal elements for the flipped coin $\tilde{\rho}_{l,c}=\mathscr{C}_{l}\left(\rho_{l,c}\right)$,
as noted by $\rho_{uu}=\left\langle \vec{u}\right|\tilde{\rho}_{l,c}\left|\vec{u}\right\rangle _{l}$.
The probabilities $\left\{ \rho_{uu}\right\} $ can be further expressed
as

\begin{equation}
\left(\begin{array}{c}
\rho_{RR}\\
\rho_{LL}\\
\rho_{UU}\\
\rho_{DD}
\end{array}\right)=\frac{1}{4}\left(\begin{array}{cccc}
1 & -2 & -2 & -2\\
1 & -2 & 2 & 2\\
1 & 2 & -2 & 2\\
1 & 2 & 2 & -2
\end{array}\right)\left(\begin{array}{c}
1\\
\zeta_{1}\\
\zeta_{2}\\
\zeta_{3}
\end{array}\right)\label{eq:rou_uu}
\end{equation}
with
\begin{align}
\zeta_{1} & =\textrm{Re}\left(\eta_{12}\right)-\textrm{Re}\left(\eta_{34}\right)\label{eq:zeta1}\\
\zeta_{2} & =\textrm{Re}\left(\eta_{13}\right)-\textrm{Re}\left(\eta_{24}\right)\label{eq:zeta2}\\
\zeta_{3} & =\textrm{Re}\left(\eta_{14}\right)-\textrm{Re}\left(\eta_{23}\right).\label{eq:zeta3}
\end{align}
Here, $\left\{ \zeta_{i}|i=1,2,3\right\} $ is named the effective
coherence, and is determined by the difference of the real part of
the non-diagonal terms of the coin density matrix of Eq. (\ref{eq:rou4*4})

In this case, the final position distribution of the walker follows
(See Appendix for detailed derivation)
\begin{equation}
P_{(x,y)}(n)=\sum_{j}\Upsilon_{j}\left(x,y,n\right)\rho_{RR}^{\frac{j+x}{2}}\rho_{LL}^{\frac{j-x}{2}}\rho_{UU}^{\frac{n-j+y}{2}}\rho_{DD}^{\frac{n-j-y}{2}},\label{eq:distribution-2D}
\end{equation}
where

\begin{equation}
\Upsilon_{j}\left(x,y,n\right)=\left(\begin{array}{c}
n\\
j
\end{array}\right)\left(\begin{array}{c}
j\\
\frac{l+x}{2}
\end{array}\right)\left(\begin{array}{c}
n-j\\
\frac{n-l+y}{2}
\end{array}\right).
\end{equation}
This is a quadrinomial distribution. Note that the summation here
has a restriction on $l$, that is, $l+x$ and $n-l+y$ must be even.
For those points $(x,y)$ not satisfied this restriction, the probability
is zero. The non-negative condition for the density matrix requires
that $\rho_{uu}$ are all non-negative, thus, there exists a limitation
for the non-diagonal terms or $\left\{ \zeta_{i}\right\} $. The allowed
values for $\left\{ \zeta_{i}\right\} $ is limited in a regular tetrahedron,
as illustrate in Fig. \ref{fig:The-limitation-on}.

\begin{widetext}

\begin{figure}
\begin{widetext}
\begin{centering}
\includegraphics[width=18cm]{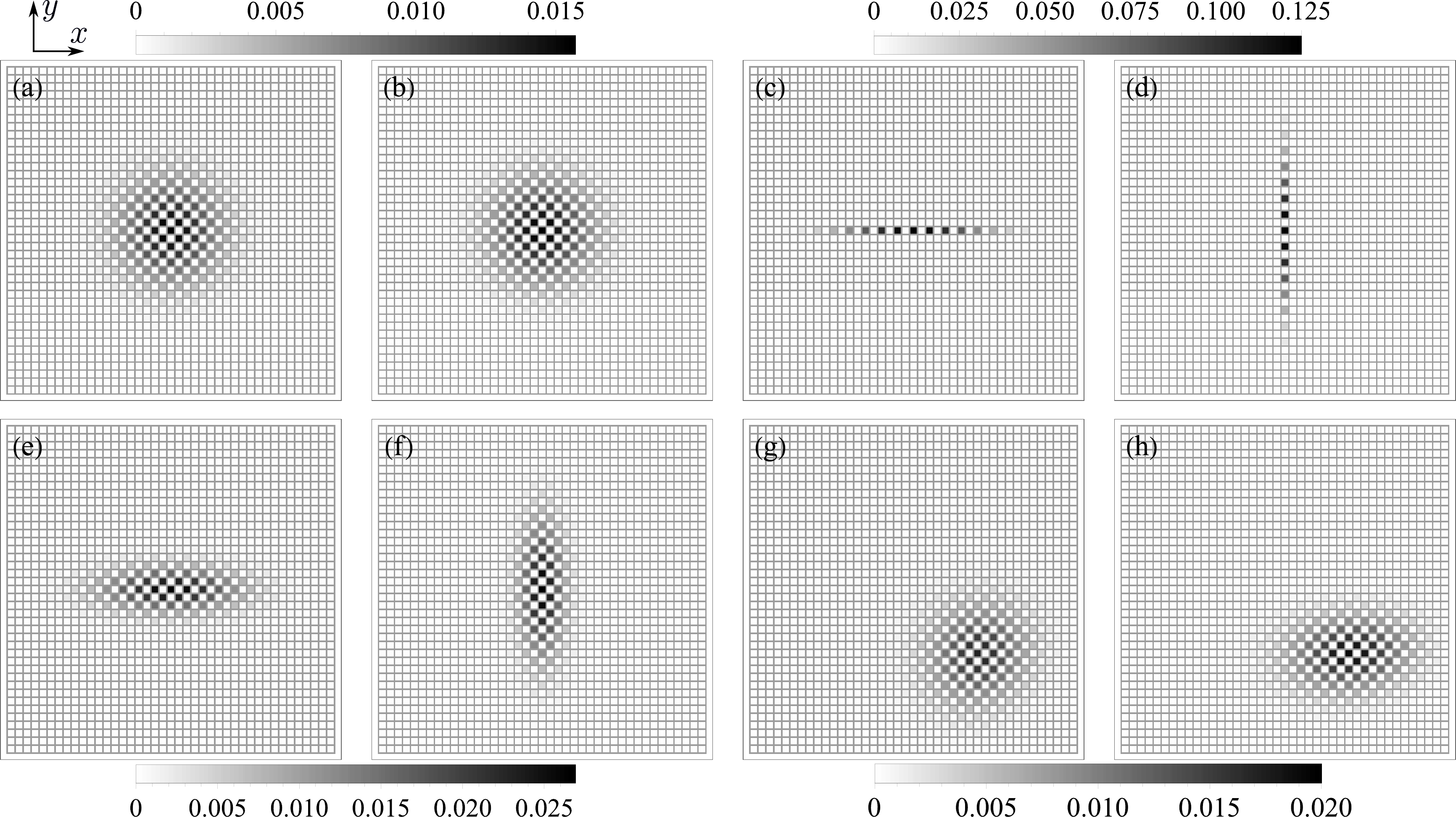}
\par\end{centering}
\caption{\label{fig:The-walker's-position}The walker's position distribution
$P_{(x,y)}(n)$ as the function of space location in quantum random
walk in 2D lattice. Here, the total step number is set as $n=40$,
and the diagonal terms in the coin's initial density matrix are chosen
as $q_{1}=q_{2}=q_{3}=q_{4}=0.25$. The subfigures are divided into
four groups, $\left\{ \textrm{(a),(b)}\right\} $, $\left\{ \textrm{(c),(d)}\right\} $,
$\left\{ \textrm{(e),(f)}\right\} $, and $\left\{ \textrm{(g),(h)}\right\} $.
In each group, the two sub-figures share the same plot-bar. (a) All
the $\eta_{ij}$ equals zero ($\zeta_{1}=\zeta_{2}=\zeta_{3}=0$).
(b) All $\eta_{ij}=0.25$, and the effect coherence diminishes ($\zeta_{1}=\zeta_{2}=\zeta_{3}=0$).
(c) Only $\eta_{12}=\eta_{21}=-\eta_{34}=-\eta_{43}=-0.25$, the others
$\eta_{ij}$ equals zero ($\zeta_{1}=-0.5,\zeta_{2}=\zeta_{3}=0$).
(d) $\eta_{12}=\eta_{21}=-\eta_{34}=-\eta_{43}=0.25$ , the others
$\eta_{ij}$ equals zero ($\zeta_{1}=0.5,\zeta_{2}=\zeta_{3}=0$).
(e) $\eta_{12}=\eta_{21}=-\eta_{34}=-\eta_{43}=-0.2$, the others
$\eta_{ij}$ equals zero ($\zeta_{1}=-0.4,\zeta_{2}=\zeta_{3}=0$).
(f) $\eta_{12}=\eta_{21}=-\eta_{34}=-\eta_{43}=0.2$, the others $\eta_{ij}$
equals zero ($\zeta_{1}=0.4,\zeta_{2}=\zeta_{3}=0$). (g) $\eta_{14}=\eta_{41}=-\eta_{23}=-\eta_{32}=-0.1$,
the others $\eta_{ij}$ equals zero ($\zeta_{3}=-0.2,\zeta_{1}=\zeta_{2}=0$).
(h) $\eta_{12}=\eta_{21}=-\eta_{34}=-\eta_{43}=-0.1$, and $\eta_{23}=\eta_{32}=0.2$,
the others $\eta_{ij}$ equals zero ($\zeta_{1}=-0.2,\zeta_{2}=0,\zeta_{3}=-0.2$).
The expectation and variance of the walker's position shown in these
sub-figures are consistent with the theoretical predictions given
by Eqs. (\ref{eq:55}), (\ref{eq:45deltax2}), and (\ref{eq:45deltay2}).}
\end{widetext}

\end{figure}

\end{widetext}

When $\zeta_{i}=0,\,i=1,2,3$, the position distribution of the walker
for QRW in 2D lattice is symmetric with $\rho_{uu}=0.25,u=R,L,U,D$,
as shown in Fig. \ref{fig:The-walker's-position}(a) and (b). We also
illustrate the walker's position distribution with different $\left\{ \eta_{ij}\right\} $
in Fig. \ref{fig:The-walker's-position}, where the various patterns
show the diverse behavior of the QRW in 2D lattice. In the simulation,
the total step number is set as $n=40$, the diagonal terms in the
coin's initial density matrix are chosen as $q_{1}=q_{2}=q_{3}=q_{4}=0.25$,
the non-diagonal term $\eta_{ij}$ is set as different values in the
eight sub-figures.
\begin{center}
\begin{figure}
\centering{}\includegraphics[width=7cm]{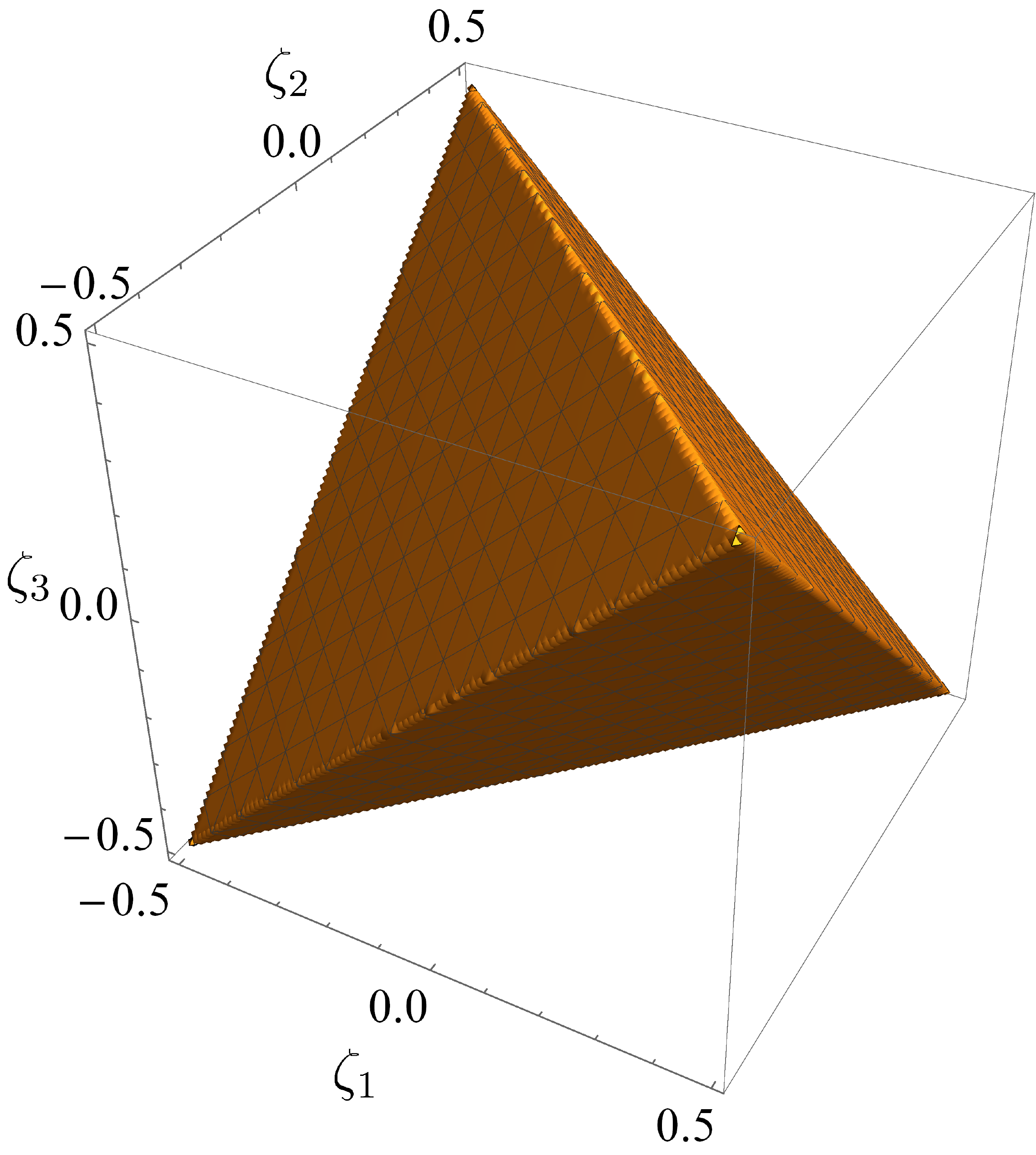}\caption{\label{fig:The-limitation-on}The limitation on $\zeta_{i}$, where
the space is expanded by $\zeta_{1}$, $\zeta_{2}$, and $\zeta_{3}$.
The allowed region for $\left\{ \zeta_{i}\right\} $ is surrounded
by the orange regular tetrahedrons obtained from the inequalities
$\rho_{RR}\protect\geq0,\,\rho_{LL}\protect\geq0,\,\rho_{UU}\protect\geq0,\,\rho_{DD}\protect\geq0$
with Eq. (\ref{eq:rou_uu}).}
\end{figure}
\par\end{center}

Then we focus on the expectation and the variance of the walker's
position. Since each step is independent for the quadrinomial distribution,
the expectation of the walker's position follows from Eq. (\ref{eq:distribution-2D})
as (detailed derivation in the Appendix \ref{subsec:2D-quantum-random})
\begin{align}
\left\langle \vec{r}\right\rangle  & =n(-\zeta_{2}-\zeta_{3},-\zeta_{2}+\zeta_{3}),\label{eq:55}
\end{align}
The variances of the walker's position along the $x$ and $y$ direction
are 
\begin{align}
\left\langle \Delta x^{2}\right\rangle  & =n[\frac{1}{2}-\zeta_{1}-(\zeta_{2}+\zeta_{3})^{2}],\label{eq:45deltax2}\\
\left\langle \Delta y^{2}\right\rangle  & =n[\frac{1}{2}+\zeta_{1}-(\zeta_{2}-\zeta_{3})^{2}],\label{eq:45deltay2}
\end{align}
respectively. The total variance for $\vec{r}$ is

\begin{align}
\left\langle \Delta\vec{r}^{2}\right\rangle  & =n\left(1-2\zeta_{2}^{2}-2\zeta_{3}^{2}\right).\label{deltar2-qrw2d}
\end{align}
These relations are cheeked by the exact numerical results illustrated
in Fig. (\ref{fig:The-walker's-position}). Equation (\ref{eq:55})
shows that only $\zeta_{2}$ and $\zeta_{3}$ determine the orientation
at $x$ or $y$, while $\zeta_{1}$ does not. According to Eq. (\ref{eq:rou_uu}),
$p_{x}=p_{LL}+p_{RR}=1/2-$$\zeta_{1}$, namely, $\zeta_{1}$ only
determines the probability that the walker moves along $x$ or $y$
direction. Different from 1D case, where the non-zero $\eta$ leads
to orientation of the walker, in 2D case, the effect of the coherence
might cancel with each other for some suitable $\eta_{ij}$ that makes
the effective coherence $\zeta_{i}=0$, as shown in Eqs. (\ref{eq:zeta1}),
(\ref{eq:zeta2}), and (\ref{eq:zeta3}). This prediction is verified
with an numerical example shown in Fig. \ref{fig:The-walker's-position}(b),
where the walker's position follows symmetric distribution with non-zero
$\eta_{ij}$. The above discovery reveals a fascinating feature of
QRW in 2D lattice: even the coherence exists in the coin's initial
state, the walker may not perform directional walking.

When the probabilities $\rho_{UU}=\rho_{DD}=0$ ($\rho_{LL}=\rho_{RR}=0$),
the walker moves only along the $x$ ($y$) direction, in which situation
the coherence satisfies $\zeta_{1}=-1/2,\,\zeta_{2}=\zeta_{3}$ (
$\zeta_{1}=1/2,\,\zeta_{2}=-\zeta_{3}$ ). The orientation is then
only determined by the effective coherence $\zeta_{3}$, and thus
the QRW in 2D lattice in this case returns to the 1D QRW, as demonstrated
in Fig. \ref{fig:The-walker's-position}(c) (Fig. \ref{fig:The-walker's-position}(d)).

\section{Conclusion and Discussion\label{sec:Conclusion-and-Discussion}}

In this paper, we extend classical random walk (CRW) to quantum random
walk (QRW) via the ensemble interpretation, and clarify the relation
between CRW, QRW, and QW (see Tab. \ref{table:The-relation-between-1}).
QRW is quantum extension of CRW from the ensemble interpretation,
while QW is the quantum extension of CRW from the single-coin interpretation.

Observed the difference of the position distribution for CRW/QRW (binomial)
and QW (ballistic), we interpret the different position distribution
from the correlation aspect. In CRW, the flipping process in each
step is independent, and thus no correlation exists between different
steps. We obtain a binomial distribution for the walker's position\citep{VANKAMPEN2007ix}.
In QRW, the walker flips different coins at different steps. Still
no correlation exists, and we retain the binomial distribution. In
QW, the sequential unitary evolution engenders strong correlation
between each steps. To qualify the correlation between different steps,
we calculate the covariance between the initial coin state and final
coin state in those walks. The result shows that the covariance is
non-zero for QW while zero for CRW/QRW.

It is found that in QRW the walker performs directional walking once
the coherence exists in the coin's initial state. We further prove
that, in such case, the stronger the coherence is, the more obvious
the directional movement is, and the smaller the fluctuation of the
walker's position distribution is. Besides, QRW in 2D lattice is also
studied, where the influence of coin state's coherence on the walker's
position distribution is found to be more complicated (than that in
the 1D case). Different from the one-dimensional case, even if there
exists coherence in the coin's initial state, the walker may not perform
directional walking. This is because, under some special conditions,
the influence of different non-diagonal terms in the coin's density
matrix on the position distribution of the walker may cancel each
other out .

Generally, the main difference of QRW and QW can be understood by
the following statement. In QRW, the quantum property refers to the
initial coherence of the coin state, which results in a directional
walk for the walker. While in QW, the sequential unitary operation
on the single coin engenders strong correlation between different
steps. This strong correlation results in the non-binomial distribution
for the walker's position.
\begin{acknowledgments}
We thank Hui Dong, Yi-Mu Du and Peng Xue for helpful suggestions for
the writing of this manuscript. This work is supported by NSFC (Grants
No. 11534002), the National Basic Research Program of China (Grant
No. 2016YFA0301201 \& No. 2014CB921403), and the NSAF (Grant No. U1730449
\& No. U1530401).
\end{acknowledgments}

\appendix

\section{The Covariance for quantum walk\label{sec:The-Covariance-for}}
\begin{widetext}
In this section, we derive the the coin's reduced density matrix and
the covariance for QW \citep{Abal_2006}. We assume the total system
is initially prepared in a pure state
\begin{equation}
\left|\psi_{\pm}(0)\right\rangle =\left|0\right\rangle _{w}\otimes\left|\pm1\right\rangle _{c},\label{eq:initial state qw}
\end{equation}
where $\left|\pm1\right\rangle _{c}$ describes the initial coin state
as $\left|1\right\rangle _{c}=\left(\begin{array}{cc}
1 & 0\end{array}\right)^{\mathrm{T}}$, and $\left|-1\right\rangle _{c}=\left(\begin{array}{cc}
0 & 1\end{array}\right)^{\mathrm{T}}$. The state after $n$ steps follows 
\begin{equation}
\left|\psi_{\pm}(n)\right\rangle =(TC)^{n}\left|\psi_{\pm}(0)\right\rangle ,\label{eq:staten}
\end{equation}
where $T$ and $C$ is given by Eq. (\ref{eq:T-onecoin}) and Eq.
(\ref{eq:C one coin}), respectively. To obtain the reduced density
matrix of the coin after $n$ steps, we first represent the initial
state in the momentum space as

\begin{equation}
\left|\psi_{\pm}(0)\right\rangle =\frac{1}{\sqrt{2\pi}}\int_{-\pi}^{\pi}\mathrm{d}k\left|k\right\rangle \otimes\left|\pm1\right\rangle _{c},\label{eq:state0}
\end{equation}
where
\begin{equation}
\left|k\right\rangle =\frac{1}{\sqrt{2\pi}}\sum_{x=-\infty}^{\infty}e^{ikx}\left|x\right\rangle _{w}.
\end{equation}
In the momentum space, the transition operator of Eq. (\ref{eq:T-onecoin})
is rewritten as 
\begin{equation}
T=e^{-ik}\otimes\left|1\right\rangle _{c}\left\langle 1\right|+e^{ik}\otimes\left|-1\right\rangle _{c}\left\langle -1\right|,
\end{equation}
and the evolution operator of one step follows 
\begin{equation}
TC=\frac{1}{\sqrt{2}}\left(\begin{array}{cc}
e^{-ik} & e^{-ik}\\
e^{ik} & -e^{ik}
\end{array}\right).\label{eq:TC}
\end{equation}
Combining Eqs. (\ref{eq:staten}), (\ref{eq:state0}), and (\ref{eq:TC}),
we obtain the state after $n$ steps
\begin{equation}
\left|\psi_{\pm}(n)\right\rangle =\frac{1}{\sqrt{2\pi}}\int\mathrm{d}k\left|k\right\rangle \otimes\left(\begin{array}{c}
\alpha_{k}^{(\pm)}(n)\\
\beta_{k}^{(\pm)}(n)
\end{array}\right),
\end{equation}
where

\begin{align}
\alpha_{k}^{(+)}(n) & =\frac{1}{2}\left(\kappa_{k}^{(+)}(n)+\frac{\cos k}{\sqrt{1+\left(\cos k\right)^{2}}}\kappa_{k}^{(-)}(n)\right),\label{eq:alphap+}\\
\alpha_{k}^{(-)}(n) & =\frac{e^{-ik}}{2\sqrt{1+\left(\cos k\right)^{2}}}\kappa_{k}^{(-)}(n),\\
\beta_{k}^{(+)}(n) & =\frac{e^{ik}}{2\sqrt{1+\left(\cos k\right)^{2}}}\kappa_{k}^{(-)}(n),\\
\beta_{k}^{(-)}(n) & =\frac{1}{2}\left(\kappa_{k}^{(+)}(n)-\frac{\cos k}{\sqrt{1+\left(\cos k\right)^{2}}}\kappa_{k}^{(-)}(n)\right),\label{eq:betap-}
\end{align}
and

\begin{equation}
\kappa_{k}^{(\pm)}(n)=e^{-in\omega_{k}}\pm\left(-1\right)^{n}e^{in\omega_{k}},\label{eq:kappap+-}
\end{equation}
with $\omega_{k}=\arcsin\left(\sin k/\sqrt{2}\right)$. Then, the
reduced density matrix of the coin after $n$ steps can be obtained
by tracing over the freedom of the walker as,

\begin{equation}
\rho_{c}^{(\pm)}(n)=\mathrm{Tr}_{\mathrm{walker}}\left(\left|\psi_{\pm}(n)\right\rangle \left\langle \psi_{\pm}(n)\right|\right)=\left(\begin{array}{cc}
\rho_{1,1}^{(\pm)}(n) & \rho_{1,-1}^{(\pm)}(n)\\
\rho_{-1,1}^{(\pm)}(n) & \rho_{-1,-1}^{(\pm)}(n)
\end{array}\right),
\end{equation}
which is further written as

\begin{equation}
\rho_{c}^{(\pm)}(n)=\left(\begin{array}{cc}
\frac{1}{2\pi}\int_{-\pi}^{\pi}\left|\alpha_{k}^{(\pm)}(n)\right|^{2}\mathrm{d}k & \frac{1}{2\pi}\int_{-\pi}^{\pi}\alpha_{k}^{(\pm)}(n)\left(\beta_{k}^{(\pm)}(n)\right)^{*}\mathrm{d}k\\
\frac{1}{2\pi}\int_{-\pi}^{\pi}\beta_{k}^{(\pm)}(n)\left(\alpha_{k}^{(\pm)}(n)\right)^{*}\mathrm{d}k & 1-\frac{1}{2\pi}\int_{-\pi}^{\pi}\left|\alpha_{p}^{(\pm)}(n)\right|^{2}\mathrm{d}k
\end{array}\right).
\end{equation}
Combining Eqs. (\ref{eq:alphap+}-\ref{eq:kappap+-}), we obtain the
explicit result for the elements of the reduced matrix as

\begin{align}
\rho_{1,1}^{(+)}(n) & =1-\frac{\sqrt{2}}{4}+\frac{\left(-1\right)^{n}}{4\pi}\int_{-\pi}^{\pi}\left(\frac{\cos\left(2\omega_{k}n\right)}{1+\left(\cos k\right)^{2}}\right)\mathrm{d}k\label{eq:rho11+}\\
\rho_{1,-1}^{(+)}(n) & =\frac{2-\sqrt{2}}{4}+\frac{\left(-1\right)^{n}}{4\pi}\int_{-\pi}^{\pi}e^{-ik}\left(\frac{i\sin(2\omega_{k}n)}{\sqrt{1+\left(\cos k\right)^{2}}}-\frac{\cos k\cos\left(2\omega_{k}n\right)}{1+\left(\cos k\right)^{2}}\right)\mathrm{d}k\\
\rho_{-1,1}^{(+)}(n) & =\frac{2-\sqrt{2}}{4}+\frac{\left(-1\right)^{n}}{4\pi}\int_{-\pi}^{\pi}e^{ik}\left(\frac{-i\sin(2\omega_{k}n)}{\sqrt{1+\left(\cos k\right)^{2}}}-\frac{\cos k\cos\left(2\omega_{k}n\right)}{1+\left(\cos k\right)^{2}}\right)\mathrm{d}k\\
\rho_{-1,-1}^{(+)}(n) & =\frac{\sqrt{2}}{4}-\frac{\left(-1\right)^{n}}{4\pi}\int_{-\pi}^{\pi}\left(\frac{\cos\left(2\omega_{k}n\right)}{1+\left(\cos k\right)^{2}}\right)\mathrm{d}k
\end{align}
and

\begin{align}
\rho_{1,1}^{(-)}(n) & =\frac{\sqrt{2}}{4}-\frac{\left(-1\right)^{n}}{4\pi}\int_{-\pi}^{\pi}\frac{\cos\left(2\omega_{k}n\right)}{\left(1+\left(\cos k\right)^{2}\right)}\mathrm{d}k\\
\rho_{1,-1}^{(-)}(n) & =-\frac{2-\sqrt{2}}{4}+\frac{\left(-1\right)^{n}}{4\pi}\int_{-\pi}^{\pi}e^{-ik}\left(-\frac{i\sin(2n\omega_{k})}{\sqrt{1+\left(\cos k\right)^{2}}}+\frac{\cos k\cos\left(2\omega_{k}n\right)}{1+\left(\cos k\right)^{2}}\right)\mathrm{d}k\\
\rho_{-1,1}^{(-)}(n) & =-\frac{2-\sqrt{2}}{4}+\frac{\left(-1\right)^{n}}{4\pi}\int_{-\pi}^{\pi}e^{ik}\left(\frac{i\sin(2n\omega_{k})}{\sqrt{1+\left(\cos k\right)^{2}}}+\frac{\cos k\cos\left(2\omega_{k}n\right)}{1+\left(\cos k\right)^{2}}\right)\mathrm{d}k\\
\rho_{-1,-1}^{(-)}(n) & =1-\frac{\sqrt{2}}{4}+\frac{\left(-1\right)^{n}}{4\pi}\int_{-\pi}^{\pi}\left(\frac{\cos\left(2\omega_{k}n\right)}{1+\left(\cos k\right)^{2}}\right)\mathrm{d}k\label{eq:rho-1-1-}
\end{align}

Then, the expectation by Eq. (\ref{convariance}) is obtained from
the reduced density matrix as
\begin{align}
\left\langle \sigma_{z}(n)\sigma_{z}(0)\right\rangle  & =p_{1}\rho_{1,1}^{(+)}(n)+p_{-1}\rho_{-1,-1}^{(-)}(n)-p_{1}\rho_{-1,-1}^{(+)}(n)-p_{-1}\rho_{1,1}^{(-)}(n)\label{eq:55-1}
\end{align}
which is explicitly written as
\begin{equation}
\left\langle \sigma_{z}(n)\sigma_{z}(0)\right\rangle =1-\frac{\sqrt{2}}{2}+\frac{\left(-1\right)^{n}}{2\pi}\int_{-\pi}^{\pi}\left(\frac{\cos\left(2\omega_{k}n\right)}{1+\left(\cos k\right)^{2}}\right)\mathrm{d}k,
\end{equation}
The expectations for $\sigma_{z}$ at the initial time and after $n$
steps are $\left\langle \sigma_{z}(0)\right\rangle =p_{1}-p_{-1}$
and 
\begin{equation}
\left\langle \sigma_{z}(n)\right\rangle =\left(p_{1}-p_{-1}\right)\left(1-\frac{\sqrt{2}}{2}+\frac{\left(-1\right)^{n}}{2\pi}\int_{-\pi}^{\pi}\left(\frac{\cos\left(2\omega_{k}n\right)}{1+\left(\cos k\right)^{2}}\right)\mathrm{d}k\right),
\end{equation}
respectively. Therefore we obtain the covariance of Eq. (\ref{eq:covariance exact})
in the main text. Specially, in the large $n$ limit, the integral
in Eqs. (\ref{eq:rho11+}-\ref{eq:rho-1-1-}) diminishes due to the
highly oscillated term $\cos\left(2\omega_{k}n\right)$ or $\sin\left(2\omega_{k}n\right)$.
Therefore, the reduced density matrix of the coin approaches to a
constant as

\begin{equation}
\lim_{n\rightarrow\infty}\rho_{c}^{(+)}(n)=\frac{1}{4}\left(\begin{array}{cc}
4-\sqrt{2} & 2-\sqrt{2}\\
2-\sqrt{2} & \sqrt{2}
\end{array}\right),\:\lim_{n\rightarrow\infty}\rho_{c}^{(-)}(n)=\frac{1}{4}\left(\begin{array}{cc}
\sqrt{2} & -2+\sqrt{2}\\
-2+\sqrt{2} & 4-\sqrt{2}
\end{array}\right).
\end{equation}
This indicates that $\left\langle \sigma_{z}(n)\sigma_{z}(0)\right\rangle $
of Eq. (\ref{eq:55-1}) is a constant in the large $n$ limit, which
explain why the covariance converges to a constant for $n\rightarrow\infty$,
as shown by Eq. (\ref{eq:covariance}) in the main text.

\begin{figure}
\includegraphics[width=8.5cm]{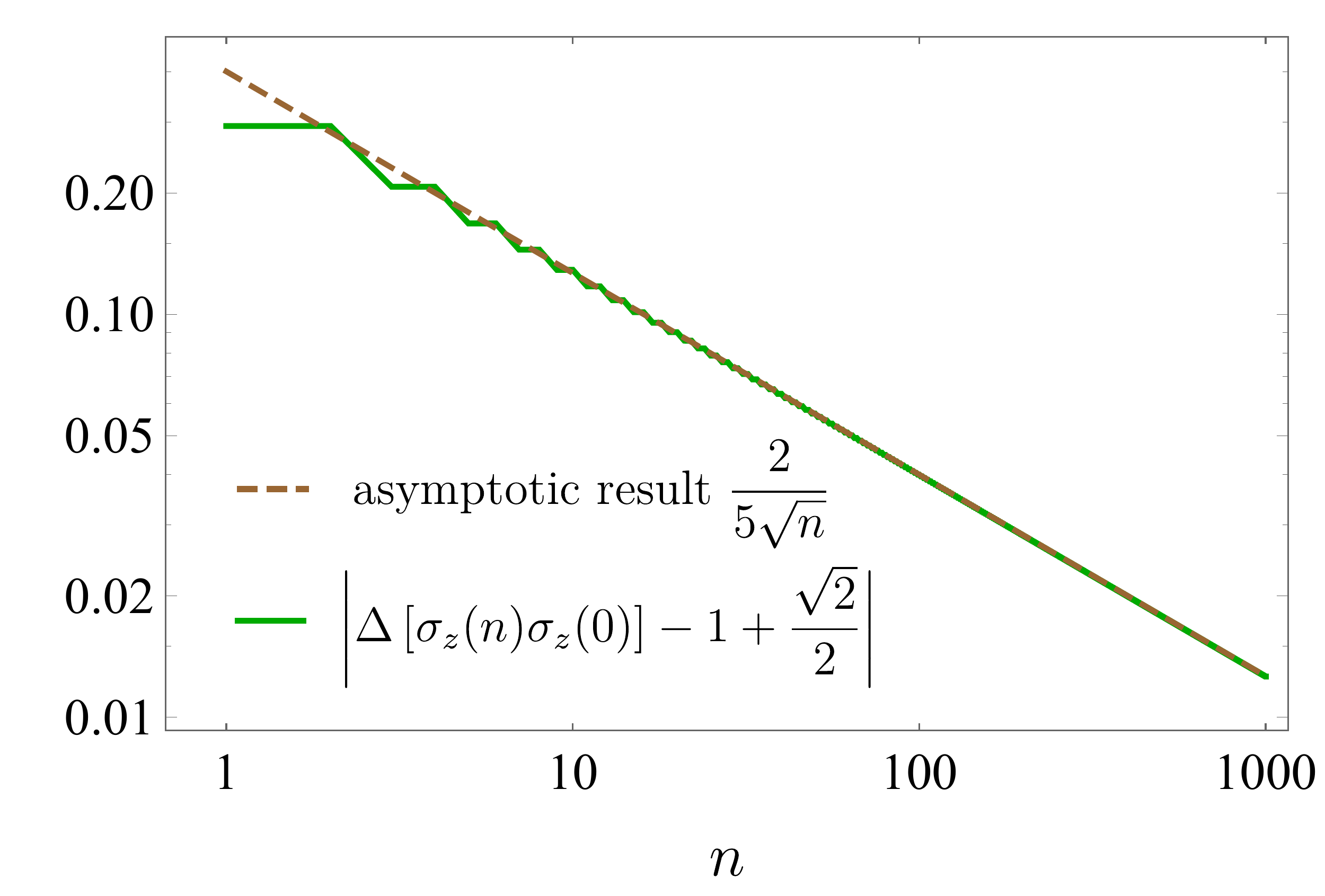}

\caption{The Log-log plot of the absolute difference of the covariance $\left\langle \Delta\left[\sigma_{z}(n)\sigma_{z}(0)\right]\right\rangle $
and the constant $1-\sqrt{2}/2$. The brown dashed line clearly shows
the absolute difference diminishes inverse proportional to $\sqrt{n}$
for large $n$. \label{fig:The-Log-log-plot}}
\end{figure}

To convince that the variance approaches to the constant $1-\sqrt{2}/2$
at long time limit, we show the absolute difference $\left|\left\langle \Delta\left[\sigma_{z}(n)\sigma_{z}(0)\right]\right\rangle -1+\sqrt{2}/2\right|$
in Fig. \ref{fig:The-Log-log-plot}. The green line clearly shows
that the absolute difference approaches to zero for long time limit
with large $n$ (the brown dashed line). By fitting the exact result
of the absolute difference, we obtain the asymptotic result (the brown
dashed line)
\begin{equation}
\left|\left\langle \Delta\left[\sigma_{z}(n)\sigma_{z}(0)\right]\right\rangle -1+\sqrt{2}/2\right|\approx\frac{2}{5\sqrt{n}},
\end{equation}
which matches well for large $n$.
\end{widetext}

\section{Quantum Random Walk in 2D Lattice}
\begin{widetext}
In this appendix, we give the detailed derivation of the walker's
position distribution and the corresponding expectation and variance
for QRW in 2D lattice. Similar to Eq. (\ref{eq:density k step}),
by acting all the coin operator first, we obtain the density matrix
after $n$ step as 
\begin{equation}
\rho(n)=\sum_{\left\{ u_{l},v_{l}\right\} }\left|\sum_{l=1}^{n}\vec{u}_{l}\right\rangle \left\langle \sum_{l=1}^{n}\vec{v}_{l}\right|\otimes\bigotimes_{l=1}^{n}\rho_{u_{l}v_{l}}\left|\vec{u}_{l}\right\rangle _{l}\left\langle \vec{v}_{l}\right|,\label{eq:72}
\end{equation}
where $u_{l}$($v_{l}$)$\in\{(1,0),\,(-1,0),\,(0,1),\,(0,-1)\}$
determines the corresponding direction $R,L,U,D$ . Tracing over the
coin Hilbert space, we obtain the probability of a given path $\left\{ u_{l}|,l=1,2,...,n\right\} $
as $P_{\left\{ u_{l}\right\} }=\prod_{l=1}^{n}\rho_{u_{l}u_{l}}$.
The probability for the walker arriving at the position $(x,y)$ after
$n$ steps is calculated with the limitation on the path 
\begin{equation}
P_{(x,y)}(n)=\sum_{\{u_{l}\}:\sum_{l}\vec{u}_{l}=(x,y)}\prod_{l=1}^{n}\rho_{u_{l}u_{l}}.
\end{equation}
If the direction $\vec{u}_{l}=(1,0),\,(-1,0),\,(0,1),\,\mathrm{and}\,(0,-1)$
is chosen for $j,\,l-j,\,m,\,$and $n-l-m$ times respectively, the
final position of the walker is $(2j-l,2m-n+l)$. The probability
for this event is quadrinomial distributed as 
\begin{equation}
P_{[j,l-j,m,n-l-m]}(n)=\left(\begin{array}{c}
n\\
l
\end{array}\right)\left(\begin{array}{c}
l\\
j
\end{array}\right)\left(\begin{array}{c}
n-l\\
m
\end{array}\right)\rho_{RR}^{j}\rho_{LL}^{l-j}\rho_{UU}^{m}\rho_{DD}^{n-l-m}.
\end{equation}
The product of the combination number $n!/[j!(l-j)!m!(n-l-m)!]$ gives
the number to divide $n$ into four group as $j,\,l-j,\,m,\,$and
$n-l-m$ . By setting the final position of the walker as $(2j-l,2m-n+l)=(x,y)$,
we can re-express $j$ and $m$ as $j=(x+l)/2$ and $m=(y+n-l)/2$
respectively. Here $j$ and $m$ need to be positive integers, which
requires the same parity for the $x$ and $y+n$. Then one can obtain
the walker's position distribution of QRW in 2D lattice as given by
Eq. (\ref{eq:distribution-2D}). The summation comes from the multiple
choice of $l$ leading to the same position $(x,y)$.

Next, we derive the expected position given in Eq. (\ref{eq:55})
and the variance of the position given in Eq. (\ref{deltar2-qrw2d}).
For the quadrinomial distribution, we can divide the final position
shift into each step for the independence of each step, and this results
in a classical probability for a summation of independent random variable
$\vec{R}(n)=\sum_{l=1}^{n}\vec{R}_{l}.$ Here, $\vec{R}_{l}=\left(X_{l},Y_{l}\right)$
is a two-component random variable which follows the independent identical
distribution and takes the value $(1,0),\,(-1,0),\,(0,1)$, and $(0,-1)$
with the probability $\rho_{RR},\,\rho_{LL},\,\rho_{UU}$, and $\rho_{DD}$
respectively. Then, the expectation for $\vec{R}_{l}$ is calculated
as

\begin{equation}
\left\langle \vec{R}_{l}\right\rangle =\rho_{RR}(1,0)+\rho_{LL}(-1,0)+\rho_{UU}(0,1)+\rho_{DD}(0,-1)=(\rho_{RR}-\rho_{LL},\rho_{UU}-\rho_{DD}),
\end{equation}
and the variances for the components $X_{l}$ and $Y_{l}$ are obtained
as

\begin{equation}
\left\langle \Delta X_{l}^{2}\right\rangle =\rho_{RR}+\rho_{LL}-(\rho_{RR}-\rho_{LL})^{2},\:\left\langle \Delta Y_{l}^{2}\right\rangle =\rho_{UU}+\rho_{DD}-(\rho_{UU}-\rho_{DD})^{2}.
\end{equation}
respectively. Thus, the variance for $\vec{R}_{l}$ follows as

\begin{equation}
\left\langle \Delta\vec{R}_{l}^{2}\right\rangle =\left\langle \Delta X_{l}^{2}\right\rangle +\left\langle \Delta Y_{l}^{2}\right\rangle =1-(\rho_{RR}-\rho_{LL})^{2}-(\rho_{UU}-\rho_{DD})^{2}.
\end{equation}
 These results can be further simplified, with the help of Eq. (\ref{eq:rou_uu}),
as

\[
\left\langle \Delta X_{l}^{2}\right\rangle =\frac{1}{2}-\zeta_{1}-(\zeta_{2}+\zeta_{3})^{2},\:\left\langle \Delta Y_{l}^{2}\right\rangle =\frac{1}{2}+\zeta_{1}-(\zeta_{2}-\zeta_{3})^{2},
\]
and

\[
\left\langle \vec{R}_{l}\right\rangle =(-\zeta_{2}-\zeta_{3},-\zeta_{2}+\zeta_{3}),\:\left\langle \Delta\vec{R}_{l}^{2}\right\rangle =1-2\zeta_{2}^{2}-2\zeta_{3}^{2}
\]
The expectation and variance for the walker's postion after $n$ steps
are thus $\left\langle \vec{r}\right\rangle =n\left\langle \vec{R}_{l}\right\rangle ,$
$\left\langle \Delta x^{2}\right\rangle =n\left\langle \Delta X_{l}^{2}\right\rangle ,$
$\left\langle \Delta r^{2}\right\rangle =n\left\langle \Delta Y_{l}^{2}\right\rangle $
and $\left\langle \Delta\vec{r}^{2}\right\rangle =n\left\langle \Delta\vec{R}_{l}^{2}\right\rangle $,
which are given explicitly as Eqs. (\ref{eq:55}) and (\ref{deltar2-qrw2d})
in the main text.
\end{widetext}

\bibliographystyle{apsrev4-1}
\bibliography{quantumwalk}

\end{document}